\documentclass[aps,prb,twocolumn,showpacs,preprintnumbers,amsmath,amssymb,superscriptaddress,notitlepage]{revtex4-2}

\usepackage[colorlinks=true,allcolors=blue]{hyperref}
\usepackage{txfonts}
\usepackage{graphicx}
\usepackage{color}
\usepackage{dcolumn}
\usepackage{bm}
\usepackage{feynmp}
\usepackage{braket}
\usepackage{ulem}

\begin{document}

\title{
Skyrmion nucleation on the surface of a topological insulator
}
\author{Daichi~Kurebayashi}
\email{d.kurebayashi@unsw.edu.au}
\affiliation{
School of Physics, The University of New South Wales, Sydney 2052, Australia
}
\affiliation{
Center for Emergent Matter Science, RIKEN, Wako 351-0198, Japan
}
\author{Oleg~A.~Tretiakov}
\email{o.tretiakov@unsw.edu.au}
\affiliation{
School of Physics, The University of New South Wales, Sydney 2052, Australia
}

\begin{abstract}
Skyrmion nucleation induced by spin-transfer torques at an interface of a topological insulator and a ferromagnetic insulator is investigated. Due to strong spin-orbit coupling on a surface of topological insulators, which enhances the effect of spin torques, efficient manipulation of skyrmions is expected, and therefore, topological insulators could provide the ideal platform to achieve high-performance skyrmionic devices.  Using micromagnetic simulations and energetics, we evaluate properties of the skyrmion nucleation on a surface of topological insulators, such as nucleation time, critical electric field, and skyrmion numbers. We show that the nucleation time is inversely proportional to the applied electric field. We also identify the Gilbert damping and temperature dependencies of the critical field. Furthermore, we analytically evaluate the effect of the Dzyaloshinskii-Moriya interaction and demonstrate that the temperature dependence can be explained by the reduction of a magnon excitation gap due to the self-energy corrections.
\end{abstract}

\maketitle

A magnetic skyrmion is a real-space topological object defined by its non-coplanar spin texture~\cite{Pokrovsky1979,Bogdanov1989,Muhlbauer2009, Yu2010, Finocchio2016, Fert2017, Gobel2021}. Because of their topological spin structure, skyrmions exhibit unique dynamics~\cite{Jiang2017,Litzius2017} and transport properties associated with a nontrivial quantum Berry phase resulting in emergent electromagnetic fields~\cite{Schulz2012, Nagaosa2013, Akosa2018}. On the one hand, a topological insulator (TI) is a momentum-space topological object characterized by its nontrivial band structure~\cite{Hasan2010, Qi2011}. Due to this topologically nontrivial structure, transport phenomena related to the momentum-space Berry curvature, such as the quantum spin Hall effect~\cite{Kane2005, Bernevig2006, Markus2007} and the quantum anomalous Hall effect~\cite{Rui2010, Chang2013, Checkelsky2014}, are realized and experimentally observed. Another consequence of the nontrivial bands is the appearance of metallic gapless surface states. The surface Dirac electrons mediate strong correlations between spin and current as their spin and momentum have a one-to-one correspondence, known as the spin-momentum locking. Although both TIs and skyrmions separately have been recent emergent topics in condensed matter physics, a combination of them could be an ideal platform to study the interplay of real- and momentum-space topology.

Recently, in a heterostructure consisting of a TI and a magnetic insulator, the ferromagnetic (FM) skyrmion formation has been observed by transport measurements~\cite{Yasuda2016,Chen2019,Li2021}. In addition to the conventional anomalous Hall effect, an extra Hall signal has been observed, which was attributed to the topological Hall effect arising from the emergent electromagnetic field of skyrmions. Beyond that, a real-space observation by a scanning transmission X-ray microscopy has been made and confirmed the formation of antiferromagnetically coupled N\'eel-type skyrmions at a TI interface with a ferrimagnet~\cite{Wu2020}.

Skyrmions on a TI surface are exhilarating not only from the viewpoint of emerging physics arising from the interplay of real- and momentum-space topologies but also due to their prospects for spintronic nanodevices~\cite{Fert2017, Gobel2021}.
It has been theoretically proposed that a skyrmion on a TI surface is accompanied by a nonzero charge density due to the chiral edge states~\cite{Nomura2010, Hurst2015, Andrikopoulos2016, Araki2016}. Because of this additional charge attached to it, a skyrmion can be manipulated by external electric fields without Ohmic losses from currents. Another mechanism has been proposed to manipulate skyrmions by utilizing spin-transfer torques, which are greatly enhanced due to the spin-momentum locking on the TI surface~\cite{Sakai2014, Kurebayashi2019}. Consequently, the dynamics of skyrmions is expected to be faster, which is also highly favorable for memory applications.

Although the skyrmion dynamics on a TI surface has been intensively investigated, the missing ingredient for successful applications of skyrmions in TIs is their nucleation studies. In conventional FMs, skyrmion nucleation has been explored~\cite{Heinrich2021}, for example, by employing geometric structures and local magnetic fluctuations, such as notches, edges, or impurity sites~\cite{Iwasaki2013b,Zhou2014, Sitte2017, Stier2017, Buttner2017, Leonov2018, Buttner2021} as well as by utilizing local injection of charge and spin currents~\cite{Tchoe2012, Sampaio2013, Hrabec2017, Finizio2019}. However, the effect of dissipation on the nucleation process, including the influence of the Gilbert damping and thermal fluctuations, has not been understood well and therefore requires further investigation. Moreover, skyrmion nucleation on a TI surface might be significantly different from the one in conventional FMs because of the spin-momentum locking. With increasing interest in skyrmionics with TIs, detailed studies on the skyrmion nucleation process are highly demanded and therefore are the subject of this paper. We investigate the general properties of skyrmion nucleation on a TI surface, such as nucleation time and critical field. We perform micromagnetic simulations at a finite temperature and demonstrate the Gilbert damping and temperature dependences of the nucleation process. To give physical understanding, we describe analytically the effect of temperature based on a self-energy renormalization. By treating the Dzyaloshinskii-Moriya interaction (DMI) as a perturbation within the random-phase approximation, we succeed in reproducing the temperature dependence obtained in our micromagnetic simulations.

{\it Stochastic magnetization dynamics model.--} Magnetization dynamics at finite temperature is analyzed with the stochastic Landau-Lifshitz-Gilbert (sLLG) equation~\cite{Garanin1997, Evans2012},
\begin{eqnarray}
\label{SLLG}
\frac{\partial\bm n}{\partial t} = -\frac{\gamma}{1 + \alpha^2} \left[\bm n\times \bm B^e + \alpha \bm n\times \left(\bm n \times \bm B^e \right) \right] + \bm T,
\end{eqnarray}
where $\gamma$ is the gyromagnetic ratio, $\alpha$ is the Gilbert damping constant, $\bm n(\bm r) = \bm M(\bm r)/ M_s$ is the normalized magnetization, $M_s$ is the saturation magnetization, $\bm B^e(\bm r) = -(1/M_s) \delta F_M/\delta \bm n(\bm r) + \bm B^{\rm th}$ is the effective magnetic field, $\bm B^{\rm th}$ is the thermal field, $F_M$ is the magnetic free energy, and $\bm T$ is the spin-transfer torque. On a surface of TIs, the magnetic free energy $F_M = F_{\rm ex} + F_{\rm ani} + F_{\rm Z} + F_{\rm DMI}$, where $F_{\rm ex} = (2J_0S^2/l_a)\int dV\, (\bm \nabla \bm n)^2$, $F_{\rm ani} = -K\int dV\, (n_z)^2$, $F_{\rm Z} = -M_s \int dV\, B_z n_z$, and $F_{\rm DMI}$ are the exchange, anisotropy, Zeeman, and DMI energies, respectively. Here $J_0$ is the exchange constant between local magnetic moments,  $l_a$ is the magnetic lattice constant, $S$ is the amplitude of the local moments, $K$ is the easy-axis anisotropy constant along $z$-axis, and $B_z$ is the magnetic field perpendicular to the film \footnote{Note that we have neglected the effect of demagnetizing field as it can be included as a small addition to an effective perpendicular magnetic anisotropy of a thin film.}. On a TI surface, electrons mediate anisotropic exchange interaction, namely the DMI, reflecting an inversion symmetry breaking. Then the DMI takes the form
$
F_{\rm DMI} = (D/\xi_l) \int\! dV [
n_z ( \partial_x  n_x + \partial_y n_y ) - n_x \partial_x n_z - n_y \partial_y  n_z ]
$,
where $\xi_l$ is the penetration length of a TI surface state into the magnetic insulator, $D = - J^2\left[ \Theta\left(|J| + E_F\right) - \Theta\left(|J| - E_F\right) \right]/(8\pi v_F)
$ is the DMI constant mediated by the surface Dirac electrons~\cite{Wakatsuki2015},
where $J$ is the {\it s-d} coupling constant between electron's spins and local moments, $v_F$ is the Fermi velocity, $E_F$ is the Fermi energy, and $\Theta(x)$ is the Heaviside function. 
The effective magnetic field is then given by
\begin{eqnarray}
\!\! B^e_{ x,y} &=& \frac{4J_0 S^2}{M_s l_a} \Delta n_{x, y} + \frac{2D}{M_s \xi_l} \partial_{x,y} n_z + B^{\rm th}_{ x,y},\\
\!\! B^e_z &=& \frac{4J_0 S^2}{M_s l_a} \Delta n_z + B_{ z} - \frac{2D}{M_s \xi_l} \bm \nabla \cdot \bm n + \frac{2K}{M_s} n_z + B^{\rm th}_z.
\end{eqnarray}
It is known that the spin-transfer torques on a TI surface that couples to local magnetic moments are substantially modified due to the strong spin-orbit coupling:
\begin{equation}
\bm T = -\frac{e\gamma}{M_s \xi_l} \left( \alpha_e \hat{x} + \beta_e \hat{y}\right)\left(\bm \nabla \cdot \bm n\right) E_x ,
\end{equation}
where 
$\alpha_e = \tau J^3{\rm sgn}(E_F) \Theta(E_F^2 - J^2)/(8\pi E_F^2)$
 and 
$\beta_e = \tau^2 J^2(E_F^2 - J^2) {\rm sgn}(E_F) \Theta(E_F^2 - J^2)/(8 \pi E_F^2)$
are dimensionless coefficients~\cite{Sakai2014, Kurebayashi2019}. 
These specific spin-transfer torques occur at the TI/magnetic insulator interface or in magnetic TI thin films.
A detailed derivation of the spin-transfer torques on the TI surface is given in the Appendix. 
Note that we neglected the effect of the spin-orbit torques (SOT) on the magnetization dynamics in Eq.~(\ref{SLLG}). On the TI surface, the SOT plays an important role in the magnetic dynamics because of the spin-momentum locking, however, for the skyrmion nucleation, which requires spatial inhomogeneity, the SOT does not give rise to any qualitative differences as the SOT are uniform contribution. Instead, the SOT under the DC current is equivalent to a static uniform in-plane magnetic field, therefore reducing the energy barrier between the uniform FM and skyrmion states and thus contributing to the reduction of the critical field.

For micromagnetic simulations, we discretize a space into a square lattice by using relations
$
\left. \partial_i \bm n (\bm r) \right|_{\bm r_j} \approx ( \bm n_{\bm j + \bm e_i} - \bm n_{\bm j - \bm e_i})/(2 l_a)
$, 
$
\left. \partial_i^2 \bm n(\bm r) \right|_{\bm r_j} \approx ( \bm n_{\bm j + \bm e_i} - 2 \bm n_{\bm j} + \bm n_{\bm j - \bm e_i})/l_a^2
$, 
and imposing periodic boundary conditions in $x$- and $y$-directions.
To numerically simulate finite temperature, the thermal field is defined by
$\bm B^{\rm th} = \bm \eta \sqrt{2\alpha k_B T/(M_s \gamma V \Delta t)}$,
where $\bm \eta$ is the random vector drawn from a standard normal distribution, $V$ is the average magnetic-ion volume, and $\Delta t$ is the simulation time-step.
As an integration scheme, we have employed the Heun's method with the time-step $\Delta t \sim 3$ fs. Since these simulations are all stochastic, all numerical data are obtained as the statistical average over 50 independent simulations. For typical parameters of a magnetic TI, we have used values estimated from the first-principle calculations and experiments: $v_F = 2.55\ \rm eV\, \AA$, $J = 0.15$ eV, $J_0 = 1.38 \times 10^{-23}$ J, $K = 7.25 \times10^{-27}$ J/$\AA^3$, $M_s = 1.16 \times 10^{4}$ J/(Tm$^3)$, and $l_a = \xi_l = 8.1$ \AA~\cite{Liu2010, Wang2015}.

\begin{figure}[tbp]
	\centering
	\includegraphics[width = 1\linewidth]{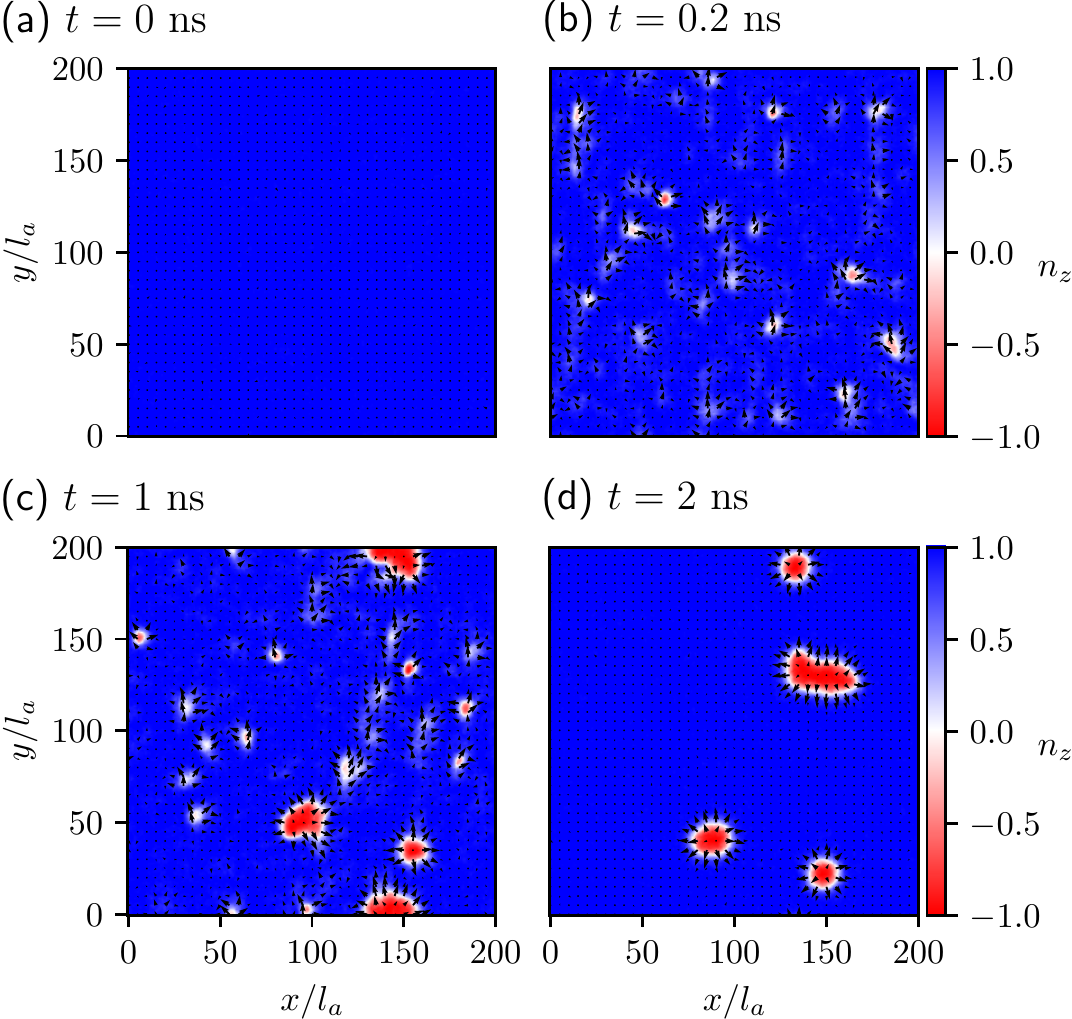}
	\caption{Magnetic profile at (a) $t = 0$ s when no current is applied, (b) $t = 0.2$ ns after current pulse is applied, (c) $t = 1$ ns after current pulse is applied, and (d) after the current pulse is switched off. The color code and arrows show the $z$- and in-plane components of magnetization, respectively. The parameters are $\alpha = 0.04$, $T = 0.2$K, and $E_x = 6.5\times 10^4$ V/m.}
	\label{fig1}
\end{figure}

{\it Skyrmion nucleation.--} We first examine the skyrmion nucleation with uniform currents. Figure~\ref{fig1} shows the magnetic profiles under a uniform current pulse.
Here, we chose the parameters as $\alpha = 0.04$, $T = 0.2$ K, and $E_x = 6.5\times 10^4$ V/m. At $t = 0$ ns, when no current is applied, the magnetization is along the $z$-axis. After application of the current pulse, it first develops particle-like small fluctuations as shown in Fig.~\ref{fig1}(b). Eventually, these particle-like fluctuations grow into N\'eel skyrmions once their radius reaches the critical one determined by $J$, $D$, and $B_z$~\cite{Bogdanov1994, Bessarab2019}, otherwise they collapse back into a uniform state. Note that, during an early stage of the nucleation, a skyrmion - antiskyrmion pair is created due to the topological number conservation~\cite{Tretiakov2007, Nagaosa2013}. However, because our DMI stabilizes only skyrmions, the antiskyrmions quickly decay into the uniform state~\cite{Potkina2020}. After the current pulse is switched off, only the skyrmions survive, see Fig.~\ref{fig1} (d). These results clearly show that skyrmions can be nucleated by uniform current pulses at TI/FM interfaces \cite{SupplMovies}.

\begin{figure}[tbp]
	\centering
	\includegraphics[width = 1\linewidth]{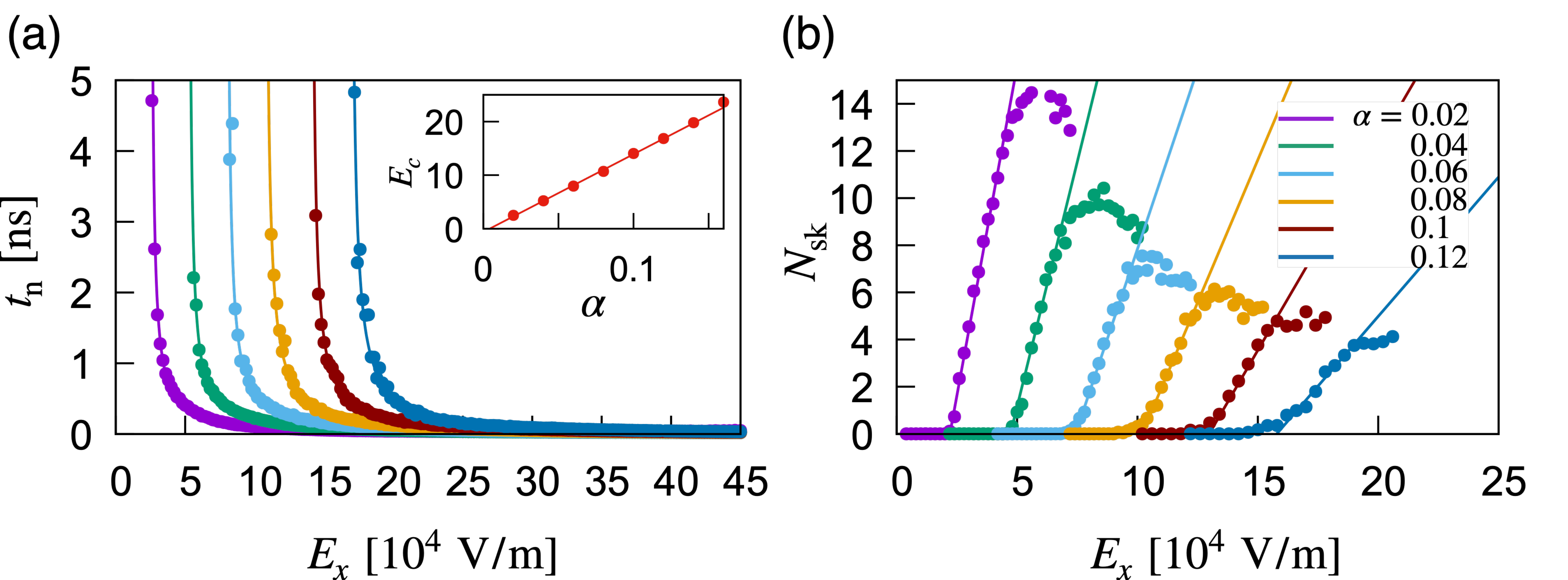}
	\caption{(a) The nucleation time and (b) the skyrmion number as a function of electric field $E_x$ are plotted for various Gilbert damping $\alpha$ and $T = 1$ K.
	The dots are numerical data and solid lines are fitting functions.	
	The inset of panel (a) shows the Gilbert damping dependence of the critical field $E_c$ in the units of $10^4$ V/m.}
	\label{fig2}
\end{figure}

{\it Effect of Gilbert damping.--} There is always a delay before the first skyrmion is nucleated. To investigate this nucleation time, we first study its Gilbert damping $\alpha$ dependence. Since $\alpha$ depends on various factors such as disorder, it is important to understand how it affects the nucleation process. The skyrmion nucleation time $t_{\rm n}$  for various $\alpha$ is shown in Fig.~\ref{fig2} (a) as a function of applied field $E_x$. We define the $t_{\rm n}$ as a time before the total skyrmion number, $N_{\rm sk} = \left|\iint dxdy\ \bm n\cdot (\partial_x \bm n \times \partial_y \bm n) \right|/(4\pi^2),$ exceeds one. The skyrmion nucleation is absent for small $E_x$, i.e., there is a critical field $E_c$ for the nucleation process. In this regime, the energy dissipation caused by the Gilbert damping exceeds the energy influx due to the spin-transfer torque \footnote{The spin-transfer torque is derived as a response to an electric field $E$, instead of current $j$. In order to determine the energy efficiency, it is required to estimate the current density, which depends on sample quality and detailed material parameters. Therefore, a theoretical estimation of energy efficiency is not straightforward. However, in general, surface states of TI are known to be less affected by impurity scattering and possess higher mobility than those of conventional metals. Consequently, the energy loss from the Joule heating is expected to be smaller.}, such that the total accumulated energy is insufficient to nucleate a skyrmion.

In terms of the nucleation time, Fig.~\ref{fig2} (a) shows diverging behavior at $E_c$ and monotonically decreases with $E_x$. This can be explained based on the energy considerations. Since the energy influx per unit time to the system is linearly proportional to $E_x$ as the coupling to electrons is treated within the linear response theory, the total accumulated energy is $F_{\rm tot} \propto E_x$. The nucleation rate $1/t_{\rm n}$ is, then, proportional to the energy difference between the total accumulated energy $F_{\rm tot}$ and the nucleation energy of a single skyrmion $F_{\rm sk}$, i.e., $1/t_{\rm n} \propto F_{\rm tot} - F_{\rm sk} \sim E_x - E_c$. Thus, the nucleation time scales as $t_{\rm n} \propto (E_x - E_c)^{-1}$. Indeed, the numerical data is in excellent agreement with it, see Fig.~\ref{fig2} (a), when fitted by, 
\begin{equation}
t_n (\alpha) = A [|E_x| - E_c(\alpha)]^{-1},
\label{eq_fit}
\end{equation}
where $A$ is a coefficient and $E_c(\alpha)$ is the critical field at the given $\alpha$. As shown in the inset of Fig.~\ref{fig2} (a), the critical field $E_c$ is linear in $\alpha$. This is because the energy dissipation is linear in $\alpha$ and the energy influx is $\propto E_x$. Thus, as the total accumulated energy is determined as the difference of the energy influx due to spin-transfer torques and the Gilbert dissipation, the required energy influx to nucleate a skyrmion should linearly increase with $\alpha$.

\begin{figure}[tbp]
	\centering
	\includegraphics[width = 1\linewidth]{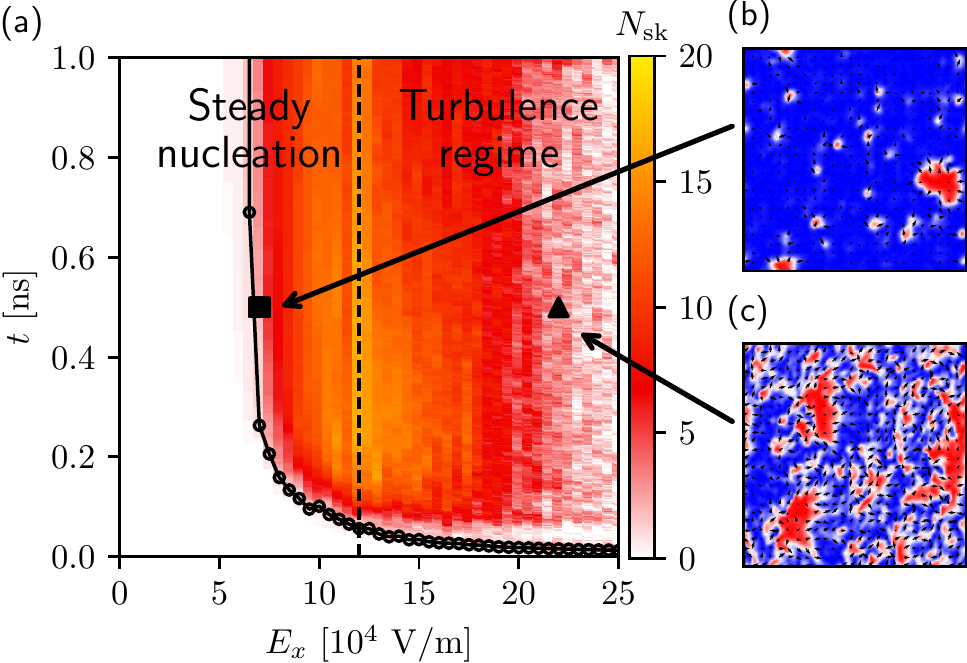}
	\caption{(a) The skyrmion number $N_{\rm sk}$ as a function of time and electric field $E_x$, for $\alpha = 0.04$ and  $T = 0.2$ K.
	(b), (c) Magnetic profile at $t = 0.5$ ns and $E_x = 7\times 10^4$ V/m, and (c) $t = 0.5$ ns and $E_x = 22\times 	10^4$ V/m.}
	\label{fig3}
\end{figure}

We also examined the total skyrmion number nucleated after 1 ns pulse as a function of applied field $E_x$ for several $\alpha$. As shown in Fig.~\ref{fig2} (b), the nucleated skyrmion number $N_{\rm sk}$ linearly increases with $E_x$ in the vicinity of the critical field. However, as $E_x$ increases further, $N_{\rm sk}$ deviates from a linear slope and saturates. The saturation occurs because skyrmions start overlapping and merging into large domains as the skyrmion density increases. By further increasing $E_x$, skyrmion states are almost destroyed by strong magnon excitations. Time evolution of $N_{\rm sk}$ at each applied field and corresponding magnetic profiles are shown in Fig.~\ref{fig3}. As seen from Fig.~\ref{fig3} (a), $N_{\rm sk}$ stays constant after reaching a steady state for the fields below $E_x \sim 1.5 \times 10^5$ m/V. In this regime, all nucleated skyrmions are well separated, as shown in Fig.~\ref{fig3} (b). On the other hand, for larger $E_x$ corresponding to the turbulence regime in Fig.~\ref{fig3} (a), $N_{\rm sk}$ decreases as $E_x$ increases and oscillates in time. Figure~\ref{fig3} (c) shows a typical magnetic profile in the strong field regime, swirling magnetic structures no longer survive \cite{SupplMovies}, and larger magnetic domains are formed because of strong magnon excitations.
Note that it is a crossover, not a phase transition, between the steady nucleation and turbulence regimes.

\begin{figure}[tbp]
	\centering
	\includegraphics[width = 1\linewidth]{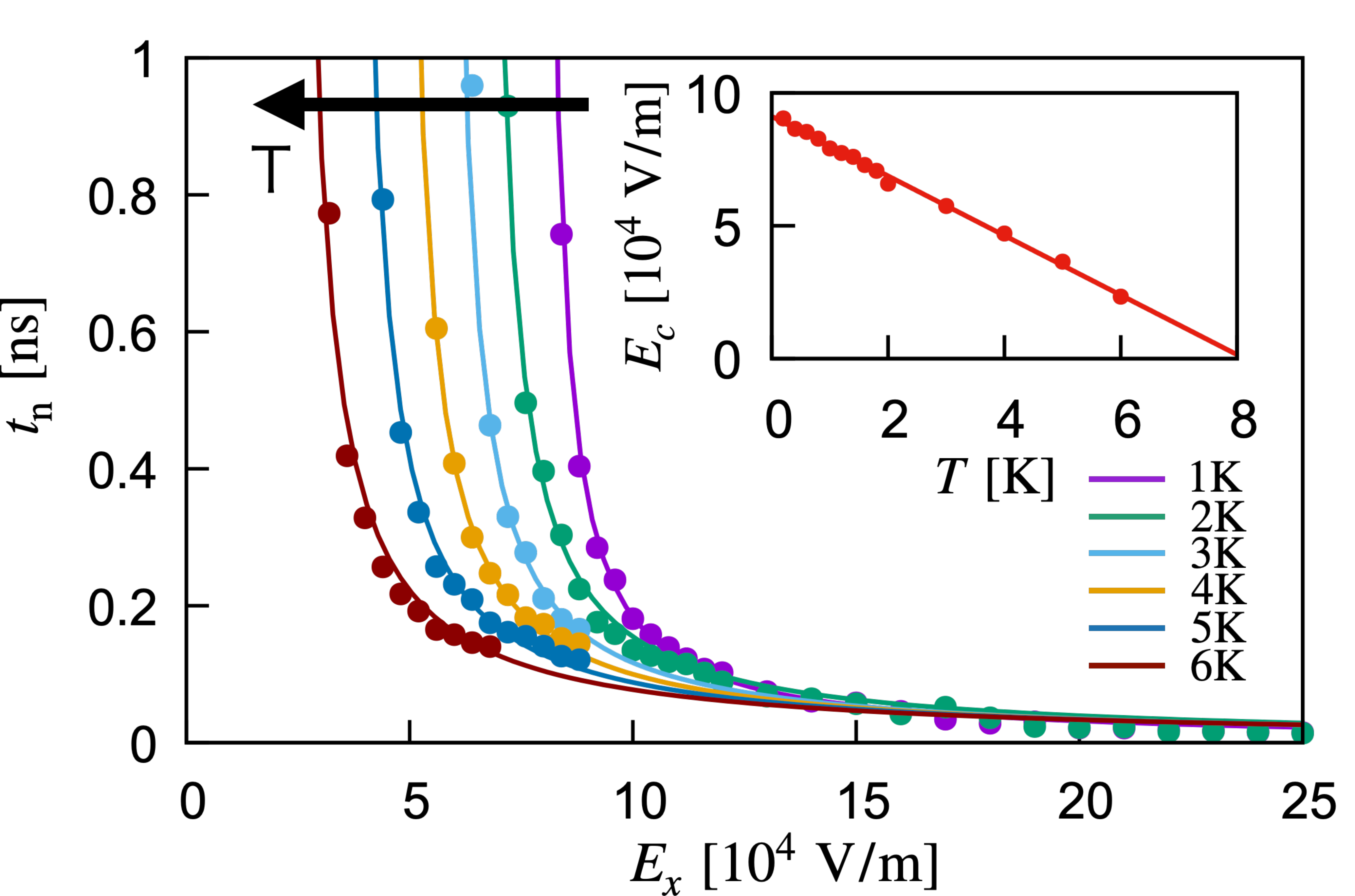}
	\caption{The nucleation time as a function of electric field $E_x$ for various temperatures and $\alpha = 0.04$. 
	The dots are numerical data, while the solid curves are fitting functions given by Eq.~(\ref{eq_fit}).
	The inset shows the temperature dependence of the critical field $E_c$.}
	\label{fig4}
\end{figure}

{\it Effect of temperature.--} Next, we examine the temperature effects on the nucleation phenomenon. The nucleation time for various temperatures is presented in Fig.~\ref{fig4}. One can notice that the temperature $T$ only affects the critical field, see $E_c (\alpha)$ in Eq.~(\ref{eq_fit}), whereas the functional form of $E_x$ is hardly modified. The critical field linearly decreases with $T$, as shown in the inset of Fig.~\ref{fig4}. Phenomenologically, the linear dependence on $T$ can be understood as follows. The thermal fluctuations supply the energy $\sim k_B T$ to the system, where $k_B$ is the Boltzmann constant. Due to this additional contribution, the energy required to create a skyrmion reduces  linearly with $T$. We note that the critical field vanishes around $T \sim 8$ K; above this temperature, skyrmions are nucleated even without the spin-transfer torques due to the thermal fluctuations.

Although phenomenological energy considerations explain the temperature effect on the nucleation process, we move one step further and try to explain this phenomenon in terms of ferromagnetic magnon excitations. Introducing the Holstein-Primakoff representation, the free energy of the system can be transformed to the magnonic Hamiltonian as
\begin{eqnarray}
\nonumber
\hat{H}_m &=& \sum_k a_{\bm k}^\dagger \left(J_k + \tilde{K} + \tilde{B} - t_k\right) a_{\bm k}\\
&&+ \sum_{\bm k,\bm q} \left( D_{\bm q} a^\dagger_{\bm q} a^\dagger_{\bm k - \bm q/2} a^{}_{\bm k + \bm q/2} + D^*_{\bm q} a^\dagger_{\bm k + \bm q/2} a^{}_{\bm k - \bm q/2} a_{\bm q} \right),
\label{H_magnon}
\end{eqnarray}
where $a_{\bm k}$ is the magnon annihilation operator with the wave number $k$, $J_k = 8 J_0 S^2 \sum_{i = x,y} (1 - \cos k_j)$ is the exchange energy, $\tilde{K} = 2K l_a^3$ is the easy-axis anisotropy, $\tilde{B} = M_s B_z l_a^3$ is the Zeeman energy,
$D_{\bm k} = 2iDl_a (\sin k_x + i \sin k_y)$ is the DMI, and $t_{\bm k} = 4e E_x l_a (\beta_e \sin k_y + \alpha_e \sin k_x)$ describes the effect of the spin-transfer torque. Note that we have retained the linear in $E_x$ term and neglected the higher-order terms. The first term gives a single-particle magnon dispersion as it already has the bilinear form in magnon operators. On the other hand, the DMI in the second line of Eq.~(\ref{H_magnon}) contains three-magnon operators and has to be treated perturbatively. 

\begin{figure}[tbp]
	\centering
	\includegraphics[width = 1\linewidth]{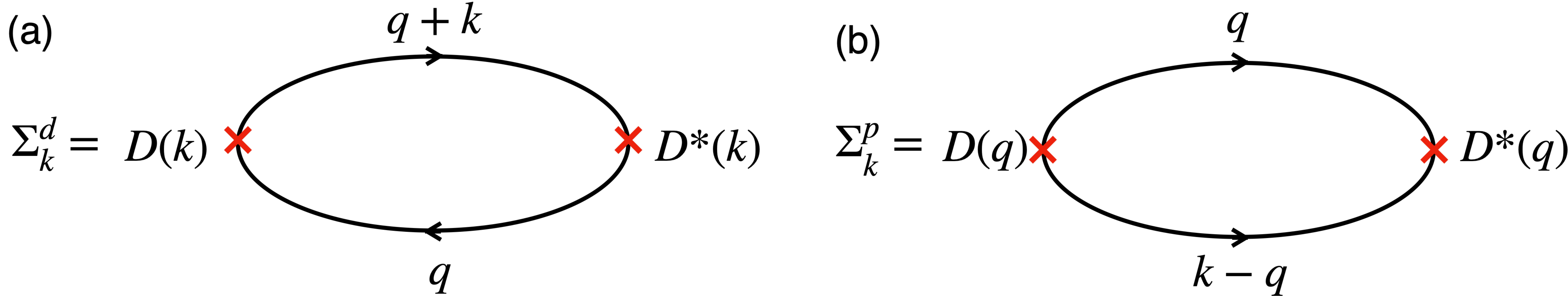}
	\caption{The Feynman diagrams contributing to the magnon self-energy. 
	(a) and (b) show the density-density and the pair correlations, respectively.}
	\label{fig5}
\end{figure}

The full Green's function is given by  $G_{\bm k}(D) = \braket{\phi_k \bar{\phi}_k} = \frac{\int D(\bar{\phi},\phi) \phi_k \bar{\phi}_k e^{-S[\phi,\bar{\phi}]}}{\int D(\bar{\phi},\phi) e^{-S[\phi,\bar{\phi}]}}$, where $S[\phi, \bar{\phi}] = \sum_k \bar{\phi}_{\bm k} (- i\omega_n) \phi_k + \mathcal{H}_m[\phi, \bar{\phi}]$ is the imaginary-time action, $\phi_{k = (\omega_n, \bm k)}$ is the eigenvalue of the magnon operator $a_{\bm k}$, and $\mathcal{H}_m$ is the Hamiltonian in the magnon coherent states basis. Within the random phase approximation, the Green's function $G_k \approx ( - i\omega_n + J_k + \tilde{K} + \tilde{B} - t_k - \Sigma_k)^{-1}$, where $\Sigma_k$ is the self-energy induced by the DMI. The real part of the self-energy modifies the magnon dispersion, while the imaginary part gives a finite lifetime. Therefore, the effective magnon dispersion including the effect of DMI takes the form
\begin{equation}
\omega_{\rm eff} (\bm k) = J_{\bm k} + \tilde{K} + \tilde{B} - t_{\bm k} - {\rm Re}\left[\Sigma_k\right].
\end{equation}
As depicted in Fig.~\ref{fig5}, the self-energy has two contributions $\Sigma_k = \Sigma_{k,d} + \Sigma_{k, p}$ with
\begin{eqnarray}
\Sigma_{k, d} &=&-4|D_{\bm k}|^2\sum_{\bm q} \frac{f_B(\omega_{\bm q}) - f_B(\omega_{\bm k + \bm q})}{\omega_{\bm q} - \omega_{\bm k + \bm q} + i 0_+}, 
\label{se_d}
\\
\Sigma_{k, p} &=& 4 \sum_{\bm q} |D_{\bm q}|^2 \frac{1 + f_B(\omega_{\bm q}) + f_B(\omega_{\bm k - \bm q})}{\omega_{\bm q} + \omega_{\bm k - \bm q} - i 0_+},
\label{se_p}
\end{eqnarray}
where $f_B$ is the Bose-Einstein distribution and $\omega_{\bm k} = J_{\bm k} + \tilde{K} + \tilde{B} - t_{\bm k}$ is the bare magnon dispersion. 
Equations~(\ref{se_d}) and~(\ref{se_p}) correspond to the magnon density-density response function, $\Sigma_d$, and pair-correlation function, $\Sigma_p$.
From Eqs.~(\ref{se_d}) and (\ref{se_p}), one notices that the real part of the self-energy is always positive, namely, the magnon correlations always reduce the magnon excitation gap min($\omega_{\rm eff}$). As the Bose-Einstein distribution function can be expanded as $f_B(\omega) \approx k_BT/\omega$, the reduction of the gap min($\omega_{\rm eff}$) due to the self-energy contributions is linear in temperature. Then, because the magnon instability is a precursor of the skyrmion nucleation and it occurs when the magnon excitation gap collapses, we conclude that the critical field linearly decreases with $T$. Note that this expansion is valid when a single-particle magnon excitation gap, $\omega_{\rm gap}$, is smaller than $k_B T$. In typical magnetically-doped TIs, the perpendicular magnetic anisotropy is $\sim 10^{-6}$ eV~\cite{Wang2015}, which corresponds to $\sim 0.1$ K. Because the condition $\omega_{\rm gap} < k_B T$ is always met within the temperature range of our simulations, the linear-$T$ dependence observed in the inset of Fig.~\ref{fig4} can be explained analytically based on the magnon excitations. Note that the instability and the nucleation of magnetic texture is a highly nonequilibrium process, and equilibrium analysis based on the self-energy corrections may fail to completely explain dynamical properties. We also neglected the magnon excitation due to the electric field, which may affect the behavior in a low-temperature regime. However, it is still capable of qualitatively describing statistical properties such as the average nucleation time and critical field.

{\it Discussion.--}
Let us review potential candidate materials for magnetic TIs.
In this study, we used material parameters for a typical magnetic TI, such as Cr- and V-doped $\rm (Bi, Sb)_2Te_3$~\cite{Liu2010, Wang2015}.
The Curie temperature of this group of materials, however, is  $T_c \approx15\sim30\rm K$, and therefore the skyrmion state is only stable at the cryogenic temperatures, thus hindering some device applications.
In order to overcome this limitation, achieving high $T_c$ is particularly desirable.
In a similar group of materials, such as $\rm MnSb_2Te_4$~\cite{Wimmer2021}, a higher Curie temperature $T_c\sim50$K is observed. Furthermore, the increase of $T_c$ by 60K due to the exchange bias effect is observed in a superlattice structure of Dy-doped $\rm Bi_2 Te_3$ and Cr-doped $\rm Sb_2Te_3$~\cite{Liu2020}. As another approach to realizing high $T_c$ magnetic TIs, bilayer systems consisting of a TI and magnetic insulator with a higher $T_c$ are expected to be more commercially realizable candidates. It is reported that a TI grown on YIG by molecular-beam epitaxy exhibits the anomalous Hall effect, i.e., a signature of magnetically-coupled surface states, up to 300K~\cite{Fanchiang2018, Pereira2020}. Thus, with recent advances in nanotechnology, there is a real prospect that room-temperature magnetic TIs will be within the reach in the near future.

In conclusion, we have comprehensively studied the current-induced skyrmion nucleation on a surface of a topological insulator. As a system, we have considered a heterostructure consisting of a TI and an ultrathin ferromagnetic insulator (applicable to the recent experimental realizations~\cite{Yasuda2016, Chen2019, Li2021, Wu2020}) and employed the specific spin-transfer torque realized on a TI surface coupling to magnetic local moments. By solving the stochastic LLG equation, we have determined the critical field $E_c$ for skyrmion nucleation and found that the nucleation time $t_n \propto (E_x - E_c)^{-1}$, where $E_c$ has been found to be proportional to the Gilbert damping $\alpha$. These results suggest that with the advances in nanotechnology, when much cleaner samples with lower damping will be available, faster skyrmion nucleation with lower fields may be achieved. Furthermore, we have investigated the temperature dependence of the nucleation and have observed the temperature effects on the critical field, which linearly decreases as temperature rises. We have given a quantum microscopic description for this linear dependence based on magnon excitations. The self-energy corrections due to the Dzyaloshinskii-Moriya interaction provide linear in temperature reduction of the magnon excitation gap and the critical field. Our results give a phenomenological understanding of the skyrmion nucleation on a surface of a topological insulator and open doors for developing TI-based skyrmionic memory and logic nanodevices.

\section*{Acknowledgments} 
The authors are grateful to N. Nagaosa for insightful discussions.
O.A.T. acknowledges the support from the Australian Research Council (Grant No.~DP200101027), the Cooperative Research Project Program at the Research Institute of Electrical Communication, Tohoku University (Japan), and the NCMAS grant.

%%%%%%%%%%%%%%%%%%%%%%%%%%%%%%%%%%%%%%%%%%%
\appendix*
\section{Spin-transfer torque on the surface of a TI}

In the appendix, we present a detailed derivation of spin-transfer torques on the surface of a topological insulator. The Hamiltonian describing the surface states of a topological insulator is given by
\begin{eqnarray}
\nonumber
H 
&=&H_0
 -e v_F \sum_{\bm k} \psi_{\bm k}^\dagger \left[\bm A(t) \times \bm \sigma\right]_z\psi_{\bm k}^{}
- J \sum_{\bm r} \bm n(\bm r) \cdot \psi_{\bm r}^\dagger \bm \sigma \psi^{}_{\bm r},\\
\end{eqnarray}
where 
$H_0 = \sum_{\bm k} \psi_{\bm k}^\dagger\left[ -v_F  \left(k_x \sigma_y - k_y \sigma_x \right) -J \sigma_z\right]\psi_{\bm k}^{}$ describes a massive Dirac Fermions with a helical spin texture~\cite{Liu2010}, 
$\psi_{\bm k}$ is the Fermionic annihilation operator at momentum $\bm k$,
$v_F$ is the Fermi velocity,
$\bm \sigma$ is the Pauli matrix describing real spin degrees of freedom,
$\bm A = (A_x, A_y)$ is the electromagnetic vector potential, 
$J$ is a {\it s-d} exchange constant, 
and $\bm n(\bm r) = \hat{\bm z} + \delta\bm n(\bm r)$ is magnetization of local moments.
Here we take the saturated magnetization direction along $\hat{\bm z}$ axis and assume $\delta \bm n(\bm r)$ is a small magnetic fluctuation around 
$\hat{\bm z}$ axis, namely $\hat{\bm z} \cdot \delta\bm n(\bm r) = 0$ and $|\delta \bm n(\bm r)| << 1$.

\begin{figure}[htbp]
	\centering
	\includegraphics[width = 0.8\linewidth]{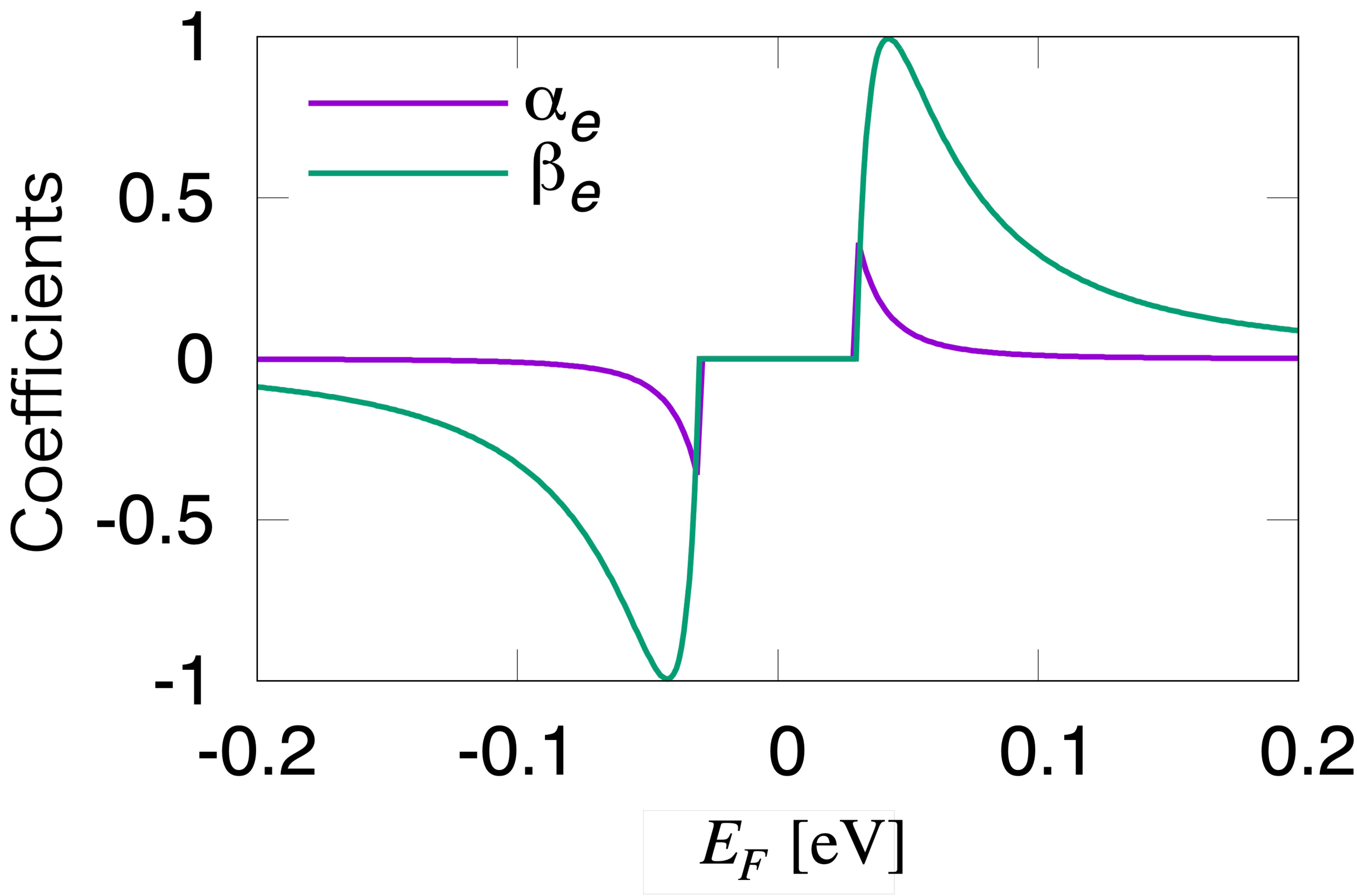}
	\caption{
		The coefficients characterizing the spin-transfer torques on the surface of the topological insulator.
		The material parameters are taken as $\frac{1}{2\tau} = \eta_0 |E_F|$, $\eta_0 = 0.1$, $v_F = 2.55\, {\rm eV\, \AA}$, and $J = 0.03$ eV. 
	}
	\label{coeff}
\end{figure}

By treating the electromagnetic field and the magnetic fluctuation, $\delta\bm n(\bm r)$, as perturbations, a non-equilibrium spin accumulation induced by an electric field applied along the $x$-direction is evaluated as
\begin{eqnarray}
	\braket{\bm \sigma(\bm r,t )} &=& -i\ {\rm Tr}\left.\left[\bm \sigma G^<(\bm r,\bm r';t,t)\right]\right|_{\bm r'\rightarrow \bm r},
	\label{spin_accm}
\end{eqnarray}
where $G^<(\bm r,\bm r;t,t) = i\braket{\psi^\dagger_{\bm r'}(t')  \psi^{}_{\bm r}(t)}$ is the lesser Green's function on the Keldysh contour and the trace is over the spin indices~\cite{tatara2008}.
Turning into momentum space, the contour ordered Green's function obeys the Dyson equation given as
\begin{eqnarray}
	\nonumber
	G_{\bm k,\bm k'} (t,t') &=& g_{\bm k}(t-t') \delta_{\bm k,\bm k'} \\
	\nonumber &&
	-ev_F \sum_{\bm k}\int_C dt_1\ g_{\bm k}(t - t_1) A_x(t_1) \sigma_y G_{\bm k,\bm k'} (t_1,t') \\
	\nonumber &&
	-J \sum_{\bm k,\bm q}\int_C dt_1\ g_{\bm k}(t - t_1)  \delta\bm n(\bm q)\cdot \bm \sigma G_{\bm k-\bm q,\bm k'} (t_1,t'),\\
\end{eqnarray}
where $G_{\bm k,\bm k'} = \sum_{\bm k,\bm k'} \exp(-i\bm k\cdot \bm r + i\bm k'\cdot \bm r' ) G(\bm r,\bm r')$ and $g_{\bm k} = [i\partial_t - H_0]^{-1}$ is an unperturbed Green's function.
By expanding the contour-ordered Green's function up to the first order in the external field and the magnetic fluctuation, $\delta\bm n(\bm r)$, one can obtain that
\begin{widetext}
	\begin{eqnarray}
		\nonumber
		G_{\bm k,\bm k'} (t,t') &\approx& -ev_F J \delta_{\bm k-\bm q,\bm k'}\sum_{\bm k,\bm q} \int_C dt_1 \int_C dt_2 \delta\bm n(\bm q)\cdot 
		\left[
			  g_{\bm k} (t-t_1) A_x(t_1)\sigma_y g_{\bm k}(t_1 - t_2) \bm \sigma g_{\bm k-\bm q}(t_2-t')
			  \right.\\
			&&\hspace{5.2cm} \left.
			+g_{\bm k + \bm q} (t-t_1)  \bm \sigma g_{\bm k}(t_1 - t_2)  A_x(t_2)\sigma_y g_{\bm k}(t_2-t')
		 \right].
	 \label{green}
	\end{eqnarray}
\end{widetext}
Note that, in Eq.~(\ref{green}), we have neglected the terms irrelevant to the non-equilibrium spin accumulation.
By substituting Eq.~(\ref{green}) into Eq.~(\ref{spin_accm}), the spin accumulation is given by
\begin{widetext}
	\begin{eqnarray}
		\nonumber \braket{\sigma_\alpha (\bm r,t )} &=&iev_F J \sum_{\bm k,\bm q}e^{i\bm q \cdot \bm r} A_x(t_1) \delta n_\beta(\bm q) {\rm Tr}\left[
		\int_C dt_1\int_C dt_2 \sigma_\alpha g_{\bm k} (t-t_1) \sigma_y g_{\bm k}(t_1 - t_2) \sigma_\beta g_{\bm k-\bm q}(t_2-t) \right.\\
		&&\hspace{3.8cm}\left.
		+\int_C dt_1\int_C dt_2 \sigma_\alpha g_{\bm k + \bm q} (t-t_2) \sigma_\beta g_{\bm k}(t_2 - t_1) \sigma_y g_{\bm k}(t_1-t)
		\right]^<.
	\end{eqnarray}
	With use of the Langreth theorem, 
	\begin{eqnarray}
		\nonumber
		\left[\int_C dt_1 \int_C dt_2 A(t-t_1)B(t_1-t_2)C(t_2-t')\right]^< &=& 
		\int_{-\infty}^\infty dt_1\int_{-\infty}^\infty dt_2 \left[A^r(t-t_1)B^r(t_1-t_2)C^<(t_2-t') \right.\\
		&&\left.
		+ A^r(t-t_1)B^<(t_1-t_2)C^a(t_2-t')
		+ A^<(t-t_1)B^a(t_1-t_2)C^a(t_2-t')\right],
	\end{eqnarray}
\end{widetext}
where superscripts $(a,r)$ denote the advanced and retarded functions, the non-equilibrium spin accumulation is obtained as
\begin{eqnarray}
	\hspace{-0.5cm}\braket{\sigma_\alpha (\bm r,t )} = \sum_{\bm q}\int \frac{d\Omega}{2\pi}e^{i\bm q \cdot \bm r - i\Omega t}\  \chi_{\alpha\beta}(\bm q,\Omega)A_x(\Omega) 	\delta n_\beta(\bm q).
\end{eqnarray}
Here $\chi_{\alpha\beta}(\bm q, \Omega)$ is a spin susceptibility defined by
\begin{widetext}
	\begin{eqnarray}
		\nonumber
		\chi_{\alpha\beta}(\bm q,\Omega) &=& iev_F J\ \int\frac{d\omega}{2\pi}\sum_{\bm k} {\rm Tr}\left[
		 \sigma_\alpha g_{\bm k} (\omega + \Omega) \sigma_y g_{\bm k}(\omega) \sigma_\beta g_{\bm k-\bm q}(\omega)
		+\sigma_\alpha g_{\bm k + \bm q} (\omega) \sigma_\beta g_{\bm k}(\omega) \sigma_y g_{\bm k}(\omega - \Omega)
		\right]^<\\
		&\approx&\frac{i\Omega}{2\pi} ev_FJ \sum_{\bm k}\left[
		 \sigma_\alpha g^r_{\bm k} \sigma_y g_{\bm k}^a \sigma_\beta g_{\bm k - \bm q}^a 
		+\sigma_\alpha g^r_{\bm k + \bm q} \sigma_\beta g_{\bm k}^r \sigma_y g_{\bm k}^a \right],
	\end{eqnarray}
\end{widetext}
where $g_{\bm k}^{r,a} \equiv g_{\bm k}^{r,a} (\omega = 0) = \left[-H_0 \pm i \eta\right]^{-1}$, $\eta$ is the damping rate induced by impurity scattering, and we have expanded the function with respect to the frequency up to the first order.
Here we have considered the small scattering regime $(\eta<<E_F,\ J)$ and only retained the dominant contributions which include both the retarded and advanced Green's functions.
By expanding the Green's function with respect to the external momentum, $\bm q$, as
$
	g_{\bm k + \bm q} \approx g_{\bm k} - v_F \epsilon_{\gamma \rho} q_\gamma g_{\bm k} \sigma_\rho g_{\bm k} + \mathcal{O}(q^2)
$,  one can obtain that
\begin{eqnarray}
	\nonumber
	\chi_{\alpha\beta}(\bm q,\Omega) 
	&\approx&
	\frac{i\Omega}{2\pi} ev_F^2J \sum_{\bm k} \epsilon_{\gamma \rho}\left[
	 	\sigma_\alpha g^r_{\bm k} \sigma_y g_{\bm k}^a \sigma_\beta g_{\bm k}^a \sigma_\rho g_{\bm k}^a
	 \right.\\
	 &&
	  \left.
		- \sigma_\alpha g^r_{\bm k} \sigma_\rho g^r_{\bm k} \sigma_\beta g_{\bm k}^r \sigma_y g_{\bm k}^a
	\right] q_\gamma.
	\label{chi}
\end{eqnarray}
Note that we have dropped the $\mathcal{O} (q^0)$-terms because they are the uniform contributions, namely corresponding to the spin-orbit torque, which is well known and there is no need to reproduce in this Appendix.
After momentum integration in Eq.~(\ref{chi}), the non-equilibrium spin accumulation corresponding to the spin-transfer torques is obtained by
\begin{eqnarray}
\braket{\bm \sigma_\perp (\bm r, t)} &=& \bm C(E_F)\bm \nabla \cdot \delta \bm n(\bm r) E_x,
\end{eqnarray}
where $\bm C(E_F) = \frac{eJ}{8\pi E_F^2}\left[ \tau^2(E_F^2 - J^2) \hat{x} - \tau J M_z \hat{y}\right] {\rm sgn(E_F)}\Theta(E_F^2 - J^2)$, 
$\tau = \frac{1}{2\eta}$ is the electron scattering lifetime,
${\rm sgn}(x)$ is the sign function, and $\Theta(x)$ is the Heaviside function.
Finally, the spin-transfer torques are given by
\begin{eqnarray}
\bm T &=& -\frac{\gamma_0 J}{M_S \xi_l} \hat{\bm z} \times \braket{\bm \sigma_\perp(\bm r,t)},
\\
T_x &=&  -\frac{e \gamma_0 \alpha_e}{M_S \xi_l} \left(\bm \nabla \cdot \delta \bm n\right)E_x,
\label{adia}
\\
T_y &=& -\frac{e\gamma_0 \beta_e}{M_S \xi_l} \left(\bm \nabla \cdot \delta \bm n\right) E_x,
\label{nonadia}
\end{eqnarray}
where $\alpha_e = \frac{\tau J^3}{8\pi E_F^2} {\rm sgn}(E_F) \Theta(E_F^2 - J^2) $ and $\beta_e = \frac{\tau^2 J^2 (E_F^2 - J^2)}{8 \pi E_F^2} {\rm sgn}(E_F) \Theta(E_F^2 - J^2)$
are dimensionless coefficients characterizing the spin-transfer torque mediated by the Dirac electrons~\cite{Sakai2014, Kurebayashi2019}.
Note that Eq.~(\ref{adia}) and~(\ref{nonadia}) correspond to the adiabatic and non-adiabatic torques, respectively.
In Fig.~\ref{coeff}, the Fermi energy dependence of the spin-transfer torque coefficients, $\alpha_e$ and $\beta_e$, are shown.
The results indicate that the non-adiabatic contribution dominates over the adiabatic one.
It is known that the non-adiabatic contribution becomes significant when the spin-relaxation is large in the system~\cite{Kohno2006}.
Due to the strong spin-orbit coupling on a surface of topological insulators, it is reasonable to have substantial non-adiabatic contributions.

\bibliography{Daichi_refs}

\end{document}